# Packaging Up Media Mix Modeling: An Introduction to Robyn's Open-Source Approach

Julian Runge [a], Igor Skokan [b], Gufeng Zhou [b], Koen Pauwels [c]

a) Northwestern University, Medill School of Journalism, Media, Integrated Marketing Communications, 1845 Sheridan Rd., Evanston, IL 60208; corresponding: julian.runge@northwestern.edu

b) Meta Platforms, Marketing Science

c) Northeastern University, D'Amore-McKim School of Business

This version: January 23, 2025

**Abstract**

As privacy-centric changes reshape the digital advertising landscape, deterministic attribution and measurement of advertising-related user behavior is increasingly constrained. In response, there has been a resurgence in the use of traditional probabilistic measurement techniques, such as media and marketing mix modeling (m/MMM), particularly among digital-first advertisers. However, small and midsize businesses often lack the resources to implement advanced proprietary modeling systems, which require specialized expertise and significant team investments. To address this gap, marketing data scientists at Meta have developed the open-source computational package Robyn, designed to facilitate the adoption of m/MMM for digital advertising measurement. This article explores the computational components and design choices that underpin Robyn, emphasizing how it "packages up" m/MMM to promote organizational acceptance and mitigate common biases. As a widely adopted and actively maintained open-source tool, Robyn is continually evolving. Consequently, the solutions described here should not be seen as definitive or conclusive but as an outline of the pathways that the Robyn community has embarked on. This article aims to provide a structured introduction to these evolving practices, encouraging feedback and dialogue to ensure that Robyn's development aligns with the needs of the broader data science community.

*Keywords*: media mix modeling, bias, prescriptive analytics, experimental calibration, computational packages, multi-objective optimization

*Acknowledgements*: The authors are grateful to the editorial and reviewer team at Harvard Data Science Review, Elea McDonnell Feit, and Brendan Sinnott for feedback and comments on earlier versions of the article.

*Competing Interests:* The first author was employed by Facebook (now Meta Platforms) until April 2021 and holds no financial interests in the company. The second and third author are employed by and hold financial interests in Meta Platforms. As open-source software, Robyn is distributed under the MIT open-source license which permits private and commercial use free of charge. All its code is publicly available and documented in detail. Neither author nor Meta Platforms obtain financial gains from Robyn.

# 1. Introduction

In the late 19th century, John Wanamaker famously said: "Half the money I spend on advertising is wasted; the trouble is I don't know which half." (Wikipedia 2025). While data availability and measurement methodologies have since improved dramatically, reliable estimates of the *incremental* effects of advertising are often still hard to come by, even in data-rich digital environments (Gordon et al. 2019; Gordon, Moakler, and Zettelmeyer 2023). Randomized control trials (RCTs) may be a remedy, can however be challenging to implement (Johnson 2023), are not always available (Eastlack and Rao 1989; Blake, Nosko, and Tadelis 2015; Lewis and Rao 2015), and are far from always used even when available (Runge, Geinitz, and Ejdemyr 2020; Runge 2020). Additionally, private and regulatory privacy initiatives are fundamentally changing the digital data landscape (Runge and Seufert 2021; Johnson, Runge, and Seufert 2022), challenging deterministic measurement approaches and bringing traditional probabilistic techniques en vogue again (Runge and Seufert 2021; Arora et al. 2023).

One such technique is media mix modeling, *mMM*, a smaller sibling of the wider-reaching marketing mix modeling, *MMM* (Arora et al. 2023). Almost as old as the term marketing mix itself that was coined by Neil Borden in 1949 (Borden 1964), MMM became a predominant analytical approach to propose strategic refinements to a company's marketing actions in pricing, promotion, product and distribution (Mela, Gupta, and Lehmann 1997; Jedidi, Mela, and Gupta 1999; Ding et al. 2020). By virtue of the historic marketing data landscape, mMM and MMM (m/MMM for short) estimate the effects of a firm's strategy choices from time series of aggregated sales and marketing mix data. However, many smaller advertisers do not have the resources to estimate and profit from the marketing mix models used by the most successful large companies. Instead, their ad budget allocation often relies on flawed methods such as deterministic decision rules, or digital attribution requiring cross-platform tracking (e.g., with third-party cookies). The quality of not requiring user-level data is a foremost reason for MMM's current renaissance as privacy initiatives alter the data infrastructure of digital advertising (Johnson, Runge, and Seufert 2022; Arora et al. 2023).

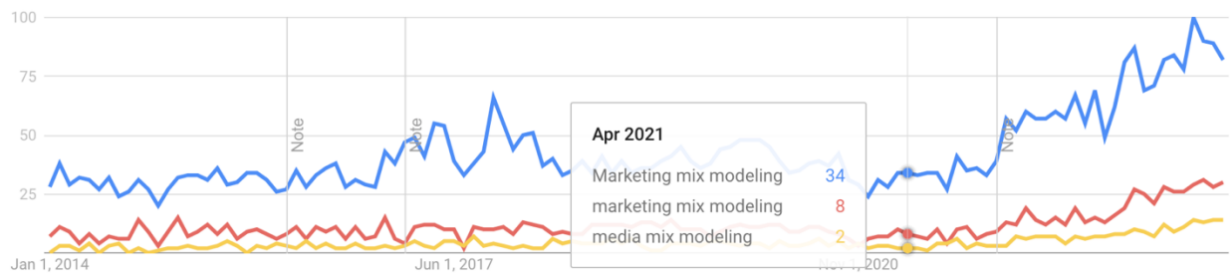

*Figure 1*: Online search interest in marketing and media mix modeling over time (Source: Google Trends 2023, worldwide). For the last two years, interest has been picking up markedly. A key factor for this development are private and public privacy initiatives, e.g., Apple's release of its App Tracking Transparency framework in April 2021; for details, see Runge and Seufert (2021).

Figure 1 shows how online search interest in m/MMM is picking up recently, a potential precipitating event being the rollout of Apple's App Tracking Transparency framework in April 2021 (Runge and Seufert 2021). Digital businesses across the board are investing into m/MMM efforts to inform how marketing spend should be allocated across different media to augment and sometimes replace prior deterministic measurement approaches (Arora et al. 2023; Runge 2023). Advertisers require guidance to not "fly blind." This issue is especially acute for small and midsize businesses that thus far relied on deterministic solutions and may not have the resources to invest in advanced proprietary modeling efforts and technology.

On this background, digital advertising platforms such as Google, Meta, and Uber have started offering solutions to support small and midsize businesses in successfully adopting mMM (Ng, Wang, and Dai 2021; CRAN 2025; Google 2023). This article delves into one of these solutions, the open-source computational package *Robyn* (Facebook Experimental 2025, 2025a). Robyn was started by marketing data scientists at Meta Platforms (CRAN 2025, CRAN 2025a) and strives to democratize access to m/MMM by supporting wide-spread successful adoption. It is distributed under the MIT open-source license which permits private and commercial use free of charge. As of January 23, 2025, it is being developed by 32 contributors, starred on Github more than 1,200 times, forked 366 times (Facebook Experimental 2025a), and has seen more than 75,000 downloads (CRAN 2025b).

As a widely adopted open-source application with an active community, Robyn is continually evolving. Consequently, the solutions described in this article should not be seen as definitive or conclusive but as an outline of the pathways that the Robyn community has embarked on. The article explains and provides context on these pathways and hopes to spark dialogue to ensure Robyn's ongoing development serves the data science community's needs.

The article proceeds by providing conceptual background on media mix modeling and laying out key computational components of Robyn vis-à-vis this conceptualization, with a focus on how Robyn supports bias mitigation and good organizational outcomes. The article then presents a detailed guide on how companies can successfully adopt Robyn, including a detailed workflow example (in Appendix A) in accordance with the offered conceptualization. It concludes by discussing limitations and key areas of future development.

## 2. Conceptual Background

### 2.1 Business problem of incrementality and choice of counterfactual

Firms want to measure the incremental impact of their media spending on outcomes (e.g., sales) to guide their future budget and allocation decision. Incremental outcomes are outcomes the firm would not have gotten without the ads, i.e. the counterfactual. This counterfactual could be regions where the ads did not run (geo-experiments), consumers who were not exposed to the ad (randomized controlled trials, RCTs), or time periods when the firm spent less, more, or no money on the ad strategy in question (m/MMM). Because the counterfactual should be as similar as possible to the actual situation, the choice in practice comes down to which comparison is most feasible and cost effective. RCTs require a randomized hold-out sample of unexposed consumers, and geo-experiments of unexposed regions, while m/MMMs require variation of ad spending in past periods of the historical data.

### *2.2. Media mix models*

We model the relevant outcome(s) as a function of the firm's marketing actions, here media spend policy in the prior period(s), and other relevant variables:

$$outcomes = f(media\ spend, other\ variables) \text{ (Eq. 1)}$$

Relevant *outcomes* could be sales, either as revenue or purchases, website visits, app installs, store visits, or a not-for-profit outcome such as donations. Relevant *other variables* are price, promotion, competitive actions, market events, seasonality, trends, and other relevant factors determining the outcome of interest (Cain 2022; Binet et al. 2023; Google 2023). The function $f$ is commonly approximated using parametric, e.g., linear, estimators (Venkatesan, Farris, and Wilcox 2021), minimizing one or several error measures, e.g., normalized root mean squared error

(NRMSE). Commonly, the input data are time series of spend by medium, i.e., advertising channel, the *other variables* described above (Binet et al. 2023), and aggregate sales or other outcomes, sometimes split out by market segments (e.g., countries, states, or DMAs).

Commonly, m/MMM assumes diminishing returns, implemented in the form of saturation curves (Johansson 1979). In other words, the model assumes that the marginal return-on-ad spend (ROAS) changes non-linearly as a response to linear spend increases. Each additional unit of media spend and advertising exposure increases the outcome, but at a declining rate.

To account for dynamic effects, modelers often include measures of *adstock* (Broadbent 1979; Leone 1995). Adstock captures the prolonged or lagged effect of advertising on demand and sales (Mela, Gupta, and Lehmann 1997). Adstock can be accounted for in models by including media spend from previous periods, often as differently weighted geometric means of ad spend in the current and prior periods (Venkatesan, Farris, and Wilcox 2021; Facebook Experimental 2025b). Some modelers also address further effects such as word-of-mouth, e.g., by including metrics on consumers' mindset or social media behavior (Srinivasan, Vanhuele and Pauwels 2010; Pauwels, Aksehirli and Lackman 2016).

## 2.2. Interpretation (and identification) of media mix models

How do we know that a model estimates the relationship between media spend-related actions and outcomes correctly? Two key issues are at play here: (1) Is there enough variation to estimate the associations between media spend and outcomes? (2) Can the estimates be interpreted as causal?

One potential source of variation that can aid in overcoming the first issue is the temporal variation in the firm's actions. As this variation in marketing allocations over time can be insufficient to generate well-identified estimates (Manchanda, Rossi, and Chintagunta 2004; Arora et al. 2023), it is important to consider intentionally amplifying it. This can be achieved by regularly implementing well-thought-out tactical changes and recording these changes in the mMM data and taking action on mMM results, i.e., changing media spend allocations. A further concern is the strategic timing of advertising that tends to intensify around high-demand holidays. Data on systematic demand shifts should hence always be included in the estimation.

Speaking to the second issue, ensuring that the estimation of Equation 1 yields causally interpretable estimates is a critical concern in media mix modeling. Observational studies of individual behavior are particularly prone to overestimating the causal impact of advertising due to "activity bias"/"user-induced endogeneity" (Lewis, Rao, and Reiley 2011). Customer-initiated ads (De Haan, Wiesel and Pauwels 2016) are triggered by visiting a particular website (page), making it tough to attribute the resulting purchase to the ad versus the consumer self-selection. In aggregate models, a similar bias arises from the strong correlation between many on- and/or offline activities, complicating the establishment of cause-effect relationships and potentially leading to systematic misestimation of advertising effects. Another common bias-introducing factor in m/MMM is confoundedness, where an unaccounted-for variable influences some of the independent variables. For instance, increased advertising activity during the holiday season may be mistakenly attributed to seasonal effects. Thus, care must be taken to include all relevant outcome drivers (Rossi 2014) – a goal post that is not perfectly attainable in practice.

To mitigate both these risks, we encourage validation of m/MMM results with experimental methods (Hanssens and Pauwels 2016; Johnson, Runge and Seufert 2022), such as field experiments (e.g., Wiesel, Pauwels and Arts 2011; Valenti et al. 2024), advertising experiments (Eastlack and Rao 1989; Blake, Nosko, and Tadelis 2015; Lewis and Rao 2015; Gordon et al. 2019; Venkatesan, Farris, and Wilcox 2021; Johnson 2023), or lower-level attribution results that reasonably reflect incremental effects (Bhargava, Galvin, and Moakler 2021). For instance, last click-attributed outcomes may approximate the causal incremental effects of campaigns, particularly when baseline effects are minimal or absent, as might occur with new or unknown products and brands where there is no organic discovery.

The challenge of identifying observational models, including m/MMMs, to provide estimates reflecting incremental effects remains significant. Integrating various approaches along the "incrementality spectrum" (Bhargava, Galvin, and Moakler 2021), especially utilizing RCTs to calibrate observational models (Runge, Patter, and Skokan 2023), is the key mitigation strategy in Robyn at this time. It was prioritized due to compelling evidence that causal inference from observational data may regularly produce misleading insights, e.g., in digital advertising (Gordon et al. 2019; Gordon, Moakler, and Zettelmeyer 2023) and pricing (Bray, Stamatopoulos, and Sanders 2024). In the future, Robyn wants to draw on methodological advances in difference-in-difference, synthetic control, instrumental variables, and panel data methods more generally to support the identification of m/MMM. We will revisit this issue in Sections 4.3 and 5.

## *2.3. The media mix modeling chain*

When used in real-world firm operations, media (and marketing) mix models are usually embedded in an analytics process that ensures that they inform the relevant actions, are understood and agreed upon by key stakeholders, and updated regularly. Figure 2 depicts an iterative five stage conceptualization of such a media mix modeling chain: The firm *acts* (1) in the marketplace with media mix policy $\pi(t)$. This policy is usually expressed as specific levels of spend allocated to available media channels and sometimes tactical options within channels. The firm then uses the observed data (in and up to period $t$) to *infer* (2) the causal effects of advertising on demand (Sanders 2019). We frame this as the estimation of a set of model candidates $M(t)$. We then include *calibration* (3) as a separate step, due to its importance per Section 2.2., to calibrate $M(t)$ with experimental or attribution results. This step can be in a "back and forth" with the inference step, e.g., calibration method and weights may change upon inspection of model candidates $M_c(t)$.

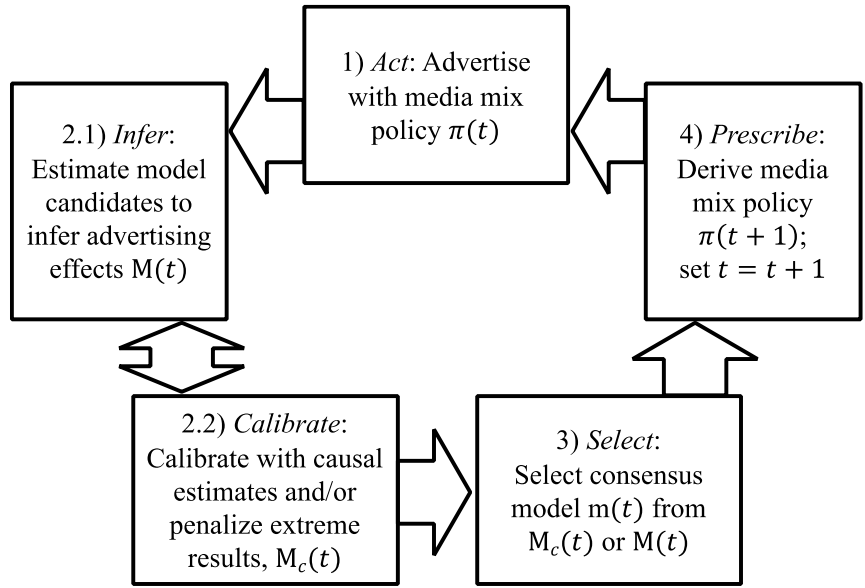

*Figure 2*: Media mix modeling (mMM) as an iterative process of the steps act, infer and calibrate, select, prescribe – and act again. Compare, e.g., to Venkatesan, Ferris, and Wilcox (2021, p. 252) or Runge, Geinitz, and Ejdemyr (2020). We list "calibrate" as a separate step to account for the calibration functionality in Robyn's multi-objective optimizer.

After producing a set of model candidates $M(t)$ or $M_c(t)$, it is incumbent upon the firm to select a consensus model $m(t)$ that it deems to accurately reflect advertising's (causal) effects on demand. This consensus model is used to prescribe a new media mix policy to be used for advertising in the next period $t + 1$. Then, the modeling chain repeats similar to the iterative marketing analytics processes described in Runge, Geinitz, and Ejdemyr (2020) or Venkatesan, Farris, and Wilcox (2021, p. 252). Period length commonly ranges from weeks to months, and sometimes quarters. We will now present computational components of Robyn against this conceptualization of the media mix modeling chain.

## 3. How Robyn Is Packaged for Organizational Acceptance and Resilience Against Biases

Biases are a common issue in business and marketing analytics. We focus on two main sources of bias: statistical and managerial bias. Statistical bias refers to the estimates of a model systematically straying from the true values. It arises from modeling mistakes, e.g., multicollinearity, endogeneity, omitted variables, unaccounted-for time series characteristics, dated and otherwise flawed data, insufficient variation and hence poor identification of the model (Venkatesan, Farris, and Wilcox 2021). Managerial bias refers to human interference with modeling (leading to statistical bias) or human error in using statistically unbiased results for decision-making. Marketing decisions often are high-stakes and involve multiple stakeholders, exacerbating these problems (Moorman 1995; Deighton, Mela, and Moorman 2021). Strong preconceptions, opinions, or badly set incentives can interfere with acceptance of the (statistically) least biased result (Moorman and Day 2016).[1]

Good marketing analytics on the organizational level require a strong tech stack (Mela and Cooper 2021), education on methods and tools, and strong organizational processes where empirical evidence can override senior opinions and preconceptions (Moorman 1995; Moorman and Day 2016), including the right talent, executive endorsement of rigorous analytics, and setting the right incentives (Mank et al. 2019; Runge 2020; Sadra 2023). Managerial acceptance and understanding are essential for analytics to become effective inside an organization. To support such acceptance, authors of computational packages can build on reasonable managerial preferences and take precautions to support good organizational outcomes. Doing so can help attenuate risks that an analytics package inadvertently leads to biased decisions or poor organizational processes.

---

[1] In addition, the dataset itself can reflect social biases inherent in current market behaviors and responses – something that is difficult to overcome through computational approaches alone. We encourage modelers to dedicate time to thinking through the potential presence of such biases in their datasets and marketing strategies. Lambrecht and Tucker (2019) present an excellent analysis of this issue and highlight avenues for remediation.

Table 1 summarizes common sources of bias and drivers of acceptance along the steps of the media mix modeling chain shown in Figure 2. It also shows how different components of Robyn aim to address sources of bias and foster organizational acceptance. Trying to remain agnostic of specific methodological paradigms, Robyn uses methods from different computational fields, including statistics, econometrics, and computer science (Proserpio et al. 2020). E.g., Robyn builds on:

- the *Nevergrad* package for fast multi-objective optimization (Facebook Research 2025),
- the *Prophet* package for time series decomposition (Facebook 2025),
- *Ridge regression* that regularizes the coefficient estimates, in this way providing an avenue to accommodate potential multicollinearity (Hilt and Seegrist 1977).

The use of Nevergrad not only makes it easy to optimize for more than one objective in the inference step, but also facilitates speedy estimation. Robyn further has built-in functionality to allow for model refresh only on newly observed data (instead of re-estimating on all data; Facebook Experimental 2025b), further increasing estimation speed and reducing the risk of working from dated data and estimates (Arora et al. 2023). Prophet has proven itself as a powerful and accurate tool for time series decomposition, capable of effectively addressing seasonal confounding factors (Taylor and Letham 2017) and Ridge regression is a much used and validated technique to deal with multicollinearity (Hilt and Seegrist 1977).

We believe the inclusion of the first two components is relatively intuitive and want to comment in more detail on the remaining six components.

|  | *Modeling step (Fig. 2)* | *Component of Robyn* | *Discussed in section* | *Step in code example (App. A)* |
|---|---|---|---|---|
| **Sources of statistical bias** | | | | |
| Model/data is outdated | *Infer* | Fast estimation with *Nevergrad* optimization package; fast re-estimation only from new data | 3.0 | Step 1 and 8 |
| Missing time series characteristics | *Infer* | Use of *Prophet* package for automated detection of trends and seasonality | 3.0 | Step 2 |
| Multi-collinearity of predictors | *Infer* | Regularization via *Ridge regression* | 3.1 | Step 3 |
| Inaccurate data or model, incl. endogeneity | *Calibrate* | Model calibration with "ground truth" results in the estimation stage through multi-objective optimization | 3.2 | Step 4 |
| **Drivers of acceptance and resilience against managerial bias** | | | | |
| Avoidance of extreme results | *Calibrate* | A preference for non-extreme results can be formalized in the calibration stage | 3.3 | Step 5 |
| Transparency and consistency | *Select* | Inference one-pager with consistent set of insights to assess model quality and select a consensus model | 3.4 | Step 5 |
| Support of good analytics processes | *Select & Prescribe* | Separation of consensus model selection from policy derivation / budget allocation | 3.5 | Step 6 |
| Tailor to organizational preferences | *Prescribe* | Different modes for prescription of a new media mix policy according to organizational preferences | 3.6 | Step 7 |

*Table 1*: Sources of bias and drivers of acceptance with respective modeling steps (per Figure 2) and components in Robyn that aim to address each. The last two columns indicate where each component is discussed in the text and where it can be controlled in the analysis workflow example shown in Appendix A.

### 3.1. Multicollinearity of predictors/features

Robyn automatically applies Ridge regression which adds a penalty term to the ordinary least squares (OLS) estimates, serving to shrink the coefficients of correlated predictors. This penalty term also reduces the variance of the coefficient estimates, making the model more stable and less sensitive to multicollinearity (Hilt and Seegrist 1977; Lewis, Rao, and Reiley 2011). Unlike other regularization techniques, notably LASSO, Ridge regression is preferred for mMM because it does not eliminate any features (by shrinking them to zero), ensuring that all variables are retained in the model. This approach helps maintain the interpretability of the model while mitigating the adverse effects of multicollinearity.

Robyn's baseline premise is that every media activity has some effect, no matter how small. Ridge regression maintains all included media variables and shrinks them proportionally to their OLS estimates (James et al. 2013). The resulting estimates should be calibrated as per the next component "model calibration."

*3.2. Model calibration*

Robyn allows users to set two optimization objectives in addition to minimization of NRMSE: avoidance of extreme results (see next component) and minimization of calibration error, i.e., mean absolute percentage error (MAPE) against estimates from experiments or vetted attribution models (Blake, Nosko, and Tadelis 2015; Lewis and Rao 2015; Gordon et al. 2019; Bhargava, Galvin, and Moakler 2021; Gordon, Moakler, and Zettelmeyer 2023).

As discussed in Section 2.2 and later in Sector 4.3, calibration serves as a method of identification that guides model estimations to align closer with a more accurate ground truth (Johnson, Runge, and Seufert 2022; Runge, Patter, and Skokan 2023). This adjustment moves the estimates up the "incrementality spectrum," bringing them closer to RCT-based causal effect estimates (Bhargava, Galvin, and Moakler 2021; Moosman and Sheridan 2023). It also addresses statistical biases (e.g., activity bias and endogeneity) that may arise from incorrect model specifications or inaccuracies related to the data, variables, or model.

New functionality included in recent releases of Robyn further allows users to set weights for these additional optimization objectives vis-à-vis the main objective of minimizing NRMSE (Zhou 2023; more on this in the next paragraph). The weight for experimental calibration is called MAPE.LIFT. Runge, Patter, and Skokan (2023) describe how experimental calibration can be used in practice and what benefits it offers.

*3.3. Avoidance of extreme results*

Firms tend to avoid extreme strategies that, e.g., strongly deviate from the status quo of insight and market understanding (Calatrava et al. 2023). Robyn allows modelers to formalize such a preference. It does so via multi-objective optimization against a second objective. This objective is minimization of a measure called Decomp.RSSD which stands for decomposition root sum of squared distances (Taylor 2023). It amounts to the distance between effect share (how many sales the model says a channel drove) and spend share (how much was spent on a channel under the prior mMM policy). Equivalent to the MAPE.LIFT minimization objective discussed in the previous section, modelers can set a weight for this optimization objective, including setting it to zero (Zhou 2023).

To the authors' best knowledge, Robyn is the first mMM computational package to apply multi-objective optimization and to allow formalizing a preference for non-extreme outcomes in the inference step. We encourage Robyn's users to build experience with this feature and its use within m/MMM analytics processes (Taylor 2023). The ability to set different weights across the objectives will be important to build familiarity with these components (Zhou 2023).

Note that, by default, the weights are set equally across objectives as (1, 1, 1) for NRMSE ("statistical error"), Decomp.RSSD ("business error"), and MAPE ("calibration error") respectively (Zhou 2023). To get started, we recommend running Robyn with optimization weights (1, 0, 0), so the "traditional" statistical approach with a single objective of minimizing NRMSE. Then try (1, 1, 0), (1, 0, 1) and (1, 1, 1) and carefully evaluate and compare results using model fit and further diagnostic outputs as shown in the inference one-pager (see next section). To reiterate: The optimization weight (1, 0, 0) will thereby implement the "standard" approach with a sole focus on minimization of NRMSE, i.e., statistical error against historical data.

*3.4. Inference one-pager*

The challenges of organizational information processing and data-driven decision-making in marketing are real and felt by advertisers every day (Moorman 1995; Moorman and Day 2016; Mank et al. 2019; Runge 2020). Transparent and easily comprehensible model output can be a catalyst of good organizational outcomes. To support users on this dimension, e.g., to circumvent the pitfalls of selective perception (Rabin and Schrag 1999), Robyn outputs a comprehensive one-pager with diagnostic statistics for each model candidate estimated in the inference step. Figure 3 shows such a one-pager as produced during a recent application of Robyn by Kumar (2023).

The one-pager summarizes key diagnostic analytics for both statistical fit (top and bottom right panels) and business plausibility. In terms of business plausibility, the one-pager allows modelers (and managers) to assess the model's contribution waterfall across all considered media (top left panel) and its implied ROI – return-on-investment, for a revenue/sales outcome – or cost-per-action – CPA, for a purchase or website/app visit outcome (two panels in the row second from top). It further depicts information on the adstock rate (left panel, second row from bottom) and immediate vs. carryover response (right panel, second row from bottom). Finally, in the bottom left panel, the one-pager displays the model's implied response curves.

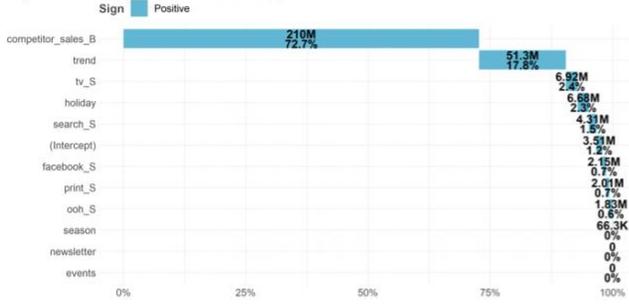
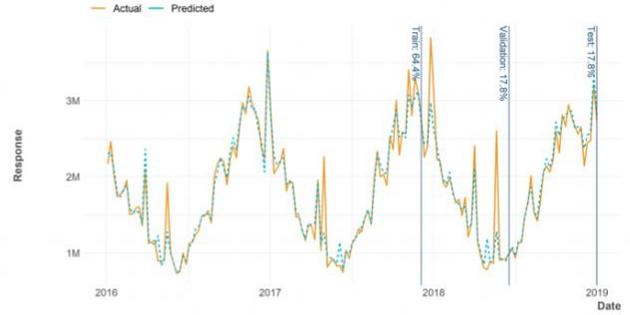
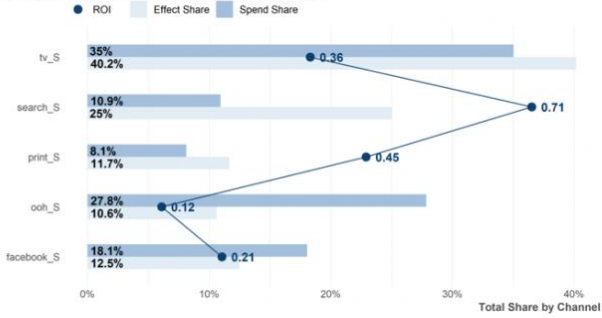
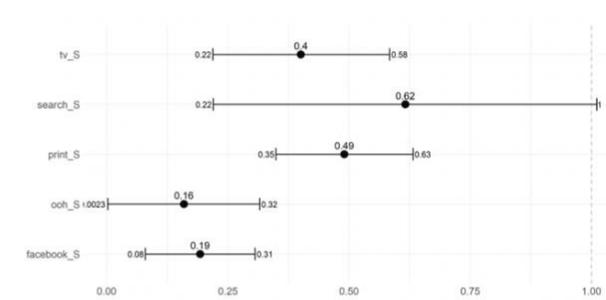
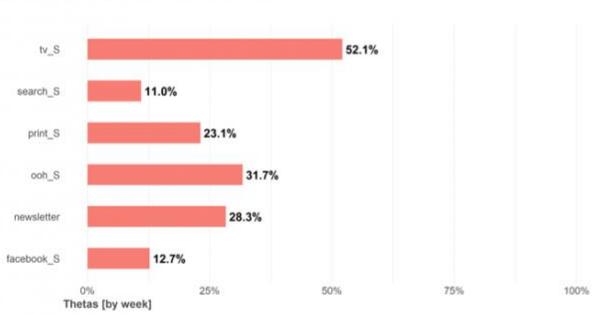
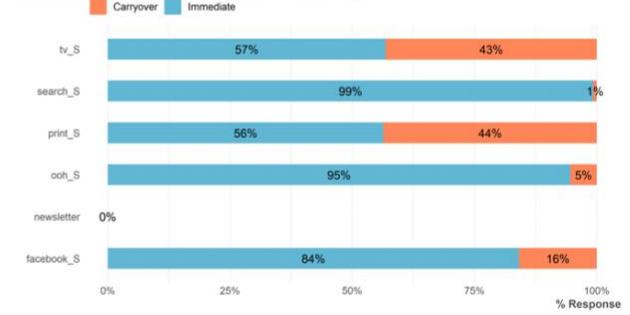
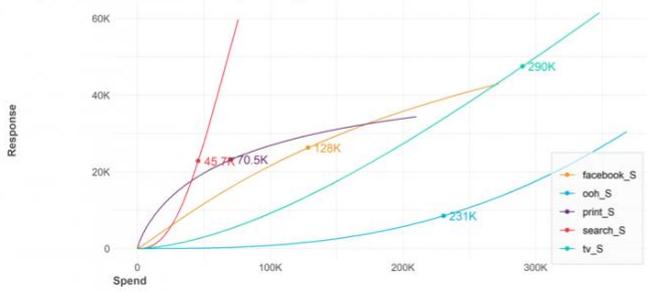
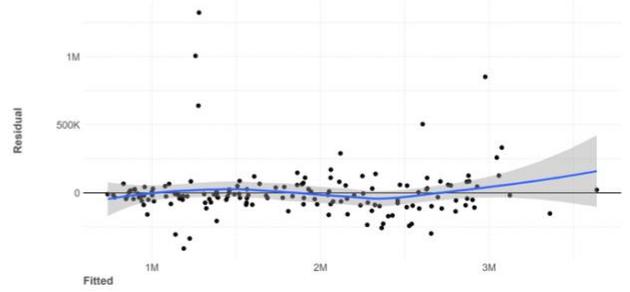

*Figure 3*: Sample inference one-pager as output by Robyn showing key diagnostic analytics for statistical fit as well as business plausibility (Source: Kumar 2023).

### 3.5. Separation of consensus model selection from prescription

The separate model selection step in our conceptualization of the media mix modeling chain and in Robyn serves an important purpose. It separates the selection of a consensus model from the inference and prescription steps. Robyn has this built-in separation to reduce the risk of managerial interference at the important and often contentious step from model results to prescription. E.g., it aims to lower the risk of authority bias to creep in, from interference by senior managers on behalf of a preferred medium or for other considerations that are not supported by rigorous analytics and data. While it may still be hard to resist the "Hippo" (highest paid person opinion) in real-world analytics operations and applied modeling (Kohavi et al. 2009), Robyn's default setup aims to support good processes and organizational outcomes: A consensus model is selected based on the results of the inference but not the prescription step.

### 3.6. Budget allocation

A way for modelers to catalyze adoption of their analytics output in the organization is to accommodate and tailor to specific organizational preferences (Moorman 1995). Aiming to support users with this, Robyn features a budget allocator that can only be run after a consensus model has been selected based on the inference one-pager. The chosen model needs to be explicitly specified in the code for the budget allocator to be able to run (Facebook Experimental 2025b; see Appendix A), supporting separation of the inference, model selection, and prescription steps per Figure 2.

In line with common approaches in m/MMM, Robyn assumes diminishing returns, implemented in the form of saturation curves (Johansson 1979). In other words, the model assumes that the marginal ROAS changes non-linearly as a response to linear spend increases. This non-linear relationship implies that there are optimal solutions for a given budget allocation problem. Robyn's budget allocator implements Augmented Lagrangian (AUGLAG) and Sequential Least Square Quadratic Programming (SLSQP) as global and local optimization algorithms drawing on the gradient-based non-linear optimization algorithm in the *nloptr* package (Facebook Experimental 2025).

Robyn currently offers two main scenarios for modelers to prescribe a new media mix policy for the next period. As discussed, these scenarios aim to help users tailor policy derivation to specific organizational preferences, they are: (1) "max_response" where Robyn is tasked with maximizing the demand response obtained with a "total_budget" and "date_range," both set by the user (Facebook Experimental 2025b). (2) "target_efficiency" that will identify the needed spend level to hit a "target_value" for ROAS for a revenue/sales outcome or CPA for a purchase or website/app visit outcome (Zhou 2022).

## 4. Successfully Adopting Robyn

In this section, we want to go from concepts to application. The Robyn Github page (Facebook Experimental 2025) has many useful resources to make it easy for analysts and data scientists to install and use Robyn. In the following, we will highlight key pitfalls and recommendations and report six community examples of companies that successfully adopted Robyn.

### *4.1. Data preparation*

A key input to successful media mix modeling is an accurate dataset where spending on each media channel varies over time. Figure 4 shows what such a dataset can look like. Customarily, an input dataset contains at least two years of weekly observations of media spend, impressions, clicks, visits, or other media-related variables, in addition to other relevant information such as important events (e.g., holidays, new product releases), competitor information (e.g., prices, sales), or weather. In direct response business models such as direct-to-consumer or digitally distributed content such as apps and games, daily data observed for shorter time periods can suffice to estimate an informative model.

Aggregating and preparing such a dataset can often require substantial investment depending on your current analytics stack and setup (Mela and Cooper 2021). Your data engineers or analysts will likely need to combine media-related information from different sources. Ideally, this assembly and checking of relevant data will be automated in dedicated data pipelines (Arora et al. 2023). They will need to vet and cross-check the data to ensure they are consistent and accurate, also over time and sometimes reaching back several years. To obtain good modeling results, they will further need to enrich the dataset with other relevant information such as competitor actions, general market trends and events, information on weather and other relevant conditions, regulatory decisions, etc. Deep and repeated conceptual reasoning is conducive to ensuring that the dataset comprehensively covers relevant information (to support you with this, consider the resources linked in Section 2.1).

Once the dataset or, better, the data pipelines have been created, we recommend performing exploratory data analysis to get a better sense of the data: What do the distributions of variables look like? Are transformations needed to address skewness, kurtosis, or other statistical properties of the data? Are there outliers? Are these there for good reasons or do they reflect errors and should be removed? Are all data recorded for all timepoints or are there missing values?

| | DATE | revenue | tv_S | ooh_S | print_S | facebook_I | search_clicks_P | search_S | competitor_sales_B | facebook_S | events | newsletter |
|---|---|---|---|---|---|---|---|---|---|---|---|---|
| 1 | 2015-11-23 | 2754371.7 | 22358.3467 | 0.000 | 12728.4889 | 24301284.2 | 0.000 | 0.000 | 8125009 | 7607.13291 | na | 19401.65 |
| 2 | 2015-11-30 | 2584276.7 | 28613.4533 | 0.000 | 0.0000 | 5527033.2 | 9837.238 | 4133.333 | 7901549 | 1141.95245 | na | 14791.00 |
| 3 | 2015-12-07 | 2547386.7 | 0.0000 | 132278.400 | 453.8667 | 16651591.2 | 12044.120 | 3786.667 | 8300197 | 4256.37538 | na | 14544.00 |
| 4 | 2015-12-14 | 2875220.0 | 83450.3067 | 0.000 | 17680.0000 | 10549765.7 | 12268.070 | 4253.333 | 8122883 | 2800.49068 | na | 2800.00 |
| 5 | 2015-12-21 | 2215953.3 | 0.0000 | 277336.000 | 0.0000 | 2934089.8 | 9467.248 | 3613.333 | 7105985 | 689.58261 | na | 15478.00 |
| 6 | 2015-12-28 | 2569921.7 | 33225.3067 | 0.000 | 31922.3111 | 16634027.0 | 12687.259 | 3773.333 | 7097237 | 5337.00517 | na | 13817.00 |
| 7 | 2016-01-04 | 2171506.7 | 687.0533 | 226753.600 | 0.0000 | 13715898.6 | 12096.505 | 3706.667 | 6974368 | 3623.01223 | na | 28817.00 |
| 8 | 2016-01-11 | 2464131.7 | 2952.2267 | 0.000 | 23581.2889 | 18851563.5 | 11588.370 | 4266.667 | 7452174 | 4129.81615 | na | 19082.00 |
| 9 | 2016-01-18 | 2012520.0 | 0.0000 | 0.000 | 0.0000 | 11910187.9 | 9341.161 | 3053.333 | 6703566 | 3623.01223 | na | 19679.00 |
| 10 | 2016-01-25 | 1738911.7 | 0.0000 | 0.000 | 11243.9111 | 10620420.2 | 7023.246 | 3053.333 | 5535394 | 3002.69059 | na | 15095.00 |
| 11 | 2016-02-01 | 1772306.7 | 16511.8267 | 42477.333 | 0.0000 | 0.0 | 9341.326 | 2746.667 | 5480098 | 0.00000 | na | 9795.00 |
| 12 | 2016-02-08 | 1809058.3 | 0.0000 | 0.000 | 4763.5556 | 11416552.5 | 7107.777 | 2693.333 | 5872523 | 3564.38220 | na | 10876.00 |
| 13 | 2016-02-15 | 1952740.0 | 66276.4400 | 0.000 | 2448.4000 | 5063253.0 | 0.000 | 0.000 | 5822926 | 1448.59774 | na | 19401.65 |
| 14 | 2016-02-22 | 1507805.0 | 0.0000 | 0.000 | 0.0000 | 0.0 | 0.000 | 0.000 | 5027123 | 0.00000 | na | 19401.65 |
| 15 | 2016-02-29 | 1510391.7 | 41556.2933 | 0.000 | 0.0000 | 10819486.9 | 5355.890 | 2293.333 | 4561518 | 2415.59640 | na | 583.00 |
| 16 | 2016-03-07 | 1588840.0 | 0.0000 | 0.000 | 16506.0889 | 0.0 | 9625.795 | 2826.667 | 4996211 | 0.00000 | na | 8224.00 |
| 17 | 2016-03-14 | 1605990.0 | 19772.4400 | 43994.133 | 0.0000 | 0.0 | 5688.412 | 2373.333 | 4938630 | 0.00000 | na | 5311.00 |
| 18 | 2016-03-21 | 1356010.0 | 0.0000 | 0.000 | 0.0000 | 0.0 | 7036.819 | 2293.333 | 4520383 | 0.00000 | na | 7428.00 |
| 19 | 2016-03-28 | 2103936.7 | 0.0000 | 66088.533 | 13579.4222 | 26747226.5 | 6090.382 | 2533.333 | 4506918 | 6471.04841 | na | 10025.00 |
| 20 | 2016-04-04 | 1120835.0 | 0.0000 | 0.000 | 0.0000 | 1146392.8 | 4636.552 | 1906.667 | 3738504 | 20.83731 | na | 9460.00 |
| 21 | 2016-04-11 | 1141140.0 | 0.0000 | 66413.867 | 0.0000 | 0.0 | 4443.257 | 1866.667 | 3768926 | 0.00000 | na | 2964.00 |
| 22 | 2016-04-18 | 1166880.0 | 11126.7333 | 0.000 | 0.0000 | 9416454.7 | 5696.908 | 1786.667 | 3663385 | 2348.68785 | na | 11663.00 |
| 23 | 2016-04-25 | 888806.7 | 0.0000 | 0.000 | 0.0000 | 0.0 | 4104.891 | 1800.000 | 2960479 | 0.00000 | na | 6317.00 |
| 24 | 2016-05-02 | 898873.3 | 0.0000 | 0.000 | 12518.5778 | 0.0 | 3685.014 | 1266.667 | 2861890 | 0.00000 | na | 5408.00 |

*Figure 4: A well-behaved sample dataset included in Robyn. As customary in media and marketing mix modeling, the dataset contains about four years of weekly observations of media spend ("_S"), impressions ("_I"), clicks ("_P"), and other relevant information such as important events (e.g., holidays, new product releases) or competitor information (e.g., prices, sales). Usually, a minimum of two years of weekly observations is recommended to estimate an m/MMM. In direct response business models such as direct-to-consumer or digitally distributed content such as apps and games, daily data observed for shorter time periods can suffice to estimate an informative model.*

## 4.2. Analysis code and workflow

Appendix A to this article includes a workflow and code example for Robyn. Table 1 shows how Robyn's key components discussed in Section 3 map onto that example. A similar but more comprehensive, well-documented, and detailed example can be found on GitHub in the file *demo.R* (Facebook Experimental 2025c). This example has been tested many times and effectively demonstrates all components, contains specific examples of inputs/outputs, and is organized as a complete workflow from the initial model build, through calibration with experiments, to model refresh and budget allocation. This example (and Robyn) includes a sample dataset if you do not (yet) have a well-behaved mMM dataset.

Additionally, one of the authors recorded a video (Zhou 2024) to guide users through the application of this code example and Meta Blueprint (2024) published an interactive online tutorial to onboard novice users with mMM and Robyn. Should you nonetheless encounter issues during application and use of Robyn, please do reach out to the active Robyn community, including the authors of this article who are able to help or triage.

*4.3. Adoption recommendations*

In our experience, the following recommendations can be helpful to avoid catastrophic failure, e.g., ill-informed decisions or unhelpful tactical changes based on Robyn:[2]

*Observe minimum data requirements*: In applied modeling, the "one in ten" rule can be a useful heuristic (Wikipedia 2025a). Robyn gives a warning if the ratio of independent variables to observations exceeds 1:10. For a model with ten independent variables, this would imply a minimum requirement of 100 observations, so two years of weekly observations. For the less-proven case of business models that experience and facilitate day-to-day changes and shifts, we recommend approximately doubling the number to half a year (180 days) of daily observations. We have seen effective models of this sort in direct-to-consumer e-commerce and game and app publishing.

*Think about and explore identification*: Does your data have variation? E.g., if you only advertised on two media, with largely constant spend levels over time, your data is likely not suited for estimation of an mMM. If you do not have meaningful variation across media and over time, consider first amplifying variation through changes in spend levels and tactics over time.

---

[2] We can give no guarantee as to the completeness of these recommendations. MMM is an iterative process and Robyn's development is ongoing. As mentioned before, when in doubt, do reach out to the Robyn community and consider hiring expert help.

*Calibrate, calibrate, calibrate*: The importance of model calibration against ground truth, using either experiments (Lewis, Rao, and Reiley 2011; Blake, Nosko, and Tadelis 2015; Feit and Berman 2019; Gordon et al. 2019; Moosman and Sheridan 2023) or established attribution models (Bhargava, Galvin, and Moakler 2020; Berman and Feit 2024) cannot be stressed enough. Runge, Patter, and Skokan (2023) provide an actionable overview what a calibration strategy can look like. Start by comparing model estimates to and selecting your consensus model based on results of advertising experiments and trusted attribution models. Then use Robyn's calibration functionality to take trusted incrementality results into account directly when building your model.

*Build organizational processes and readiness*: Figure 2 shows what a repeatable media mix modeling chain can look like. Use it to spell out your own chain and process: what people should be involved at what step, what stakeholders need to be informed when and how, with what cadence do you want to update your media mix, what meetings need to be instituted with permanence, and should there be a gated decision process with sign-off from specific people before actions are executed. Also consider educational and informational sessions that explain your modeling process and methodology to the wider organization to help build trust and acceptance.

*Vet early insights*: During the first runs of Robyn and your media mix modeling chain, conduct additional vetting of the recommended actions. E.g., during your first run, take the most reasonable and impactful recommendation, execute it, and carefully monitor if it achieves the intended effects. Consider running an experiment (ideally, an RCT, otherwise geo or switchback experiments, for additional guidance see Runge, Patter, and Skokan 2023) to evaluate the action and see if the mMM's insight was accurate.

*Make use of all model diagnostics:* Robyn's inference one-pager (see Table 1 and Section 3.4) summarizes key diagnostic information to select a consensus model. We advise particular caution against selecting a model if the plots suggest poor statistical fit (if the orange and blue line in the top right or the horizontal axis and the fitted line in the bottom right panel of Figure 3 diverge) or the response curves have an unreasonable shape. E.g., if you advertise at scale, exponential response curves for many media in the bottom left panel of Figure 3 indicate that the model may not be well suited for decision-making. Advertising response will usually be concave or S-shaped (Johansson 1979), so you should only really see exponential response curves for media that can be expected to be very far from saturation. Further, if the response decomposition waterfall in the top right panel of Figure 3 attributes most of the effect to baseline, the model is likely not set up to inform advertising strategy decisions.

### *4.4. Community examples*

In this section, we briefly report community voices from six companies that adopted Robyn for their advertising measurement. All six are small or midsize businesses that successfully used Robyn to implement a mMM and changed their advertising strategy based on insights provided by a model estimated via Robyn.

*Lemonade*, an insurtech company, combined its in-house MMM with Robyn for calibration and comparison. By leveraging a geo-testing strategy, Lemonade used Robyn's model calibration functionality to optimize its media budget allocation. As a result, Lemonade reallocated 5% of its media budget each period, often reducing the budget on certain media by up to 50%, leading to a 78% year-over-year revenue growth in the US in the last quarter of 2021 (Lemonade 2022).

*UniPegaso*, Italy's leading online university, adopted Robyn to enhance its digital advertising strategy. UniPegaso used Robyn to better understand the effectiveness of its media channels. This new approach increased incremental leads from awareness activities by 21% and reduced the cost per acquisition by 18% to 32% across different digital channels (Università Telematica Pegaso 2023).

*YOTTA by Publicis*, an agency specializing in data and technology, adopted Robyn to improve the efficiency of brand sales generation for a client. By benchmarking Robyn, YOTTA (2023) reports to have increased marketing efficiency by 10.3% and realized significant internal gains, such as up to three times faster calculations, saving up to three days of work and 20% in modeling time.

*Glint*, a company facilitating retail trading in precious metals, used Robyn with the help of its performance agency, Realtime. Robyn identified Facebook ads on Apple devices as a key driver of first purchases, leading Glint to increase its media budget for this category by 30% (Glint 2022). Additionally, Robyn highlighted the price of gold as a significant factor, accounting for 21% of first purchases, which informed Glint's media budget pacing and creative strategies.

*Bark*, a dog-centric company from New York, utilized Robyn to build its first in-house MMM analysis in just three months. This analysis enabled Bark to optimize its media budget allocation, resulting in increased overall subscription growth (Bark 2022).

*Talisa*, a US-based jewelry brand, turned to Robyn to navigate challenges in advertising strategy due to evolving data regulations. Robyn helped Talisa reallocate its media budget, leading to a 17% increase in sales in North America without an increase in overall marketing spend (Talisa 2022). The fast model updating capability of Robyn allowed Talisa to swiftly react to market changes, especially during peak seasons.

## 5. Limitations and Future Development

In this final section, we discuss limitations in the current implementation of Robyn and how future developments hope to address these and further extend the usability and scope of Robyn.

Robyn currently does not accept panel data. In the future, Robyn's architects would like to enable Robyn to accept such data, in particular across different channel and geographic groupings such that more granular recommendations on these dimensions can be provided (Aurier and Broz-Giroux 2014). Further, in Robyn's current implementation, calibration against experimental and trusted attribution results is the key mitigation strategy against model misspecification and lack of identification. Robyn's architects prioritized such calibration due to the mounting evidence that causal inference from observational data often strays far from true effects in digital advertising (Gordon et al. 2019; Gordon, Moakler, and Zettelmeyer 2023) and pricing (Bray, Stamatopoulos, and Sanders 2024). This dominant focus on calibration nonetheless is a limitation that Robyn aims to address particularly in the context of the incorporation of panel data. Panel data functionality will facilitate the use of instrumental variables, diff-in-diff, and synthetic control methods. These methods will help address issues of endogeneity and identification beyond calibration. Automated (or semi-automated) inclusion of some or all of these methods is a longer-term ambition.

Robyn's multi-objective optimization across statistical, business, and calibration error, i.e., across NRMSE, Decomp.RSSD, and MAPE.LIFT (see Sections 3.2 and 3.3), is novel in the realm of m/MMM. Modelers can flexibly set the weights for this optimization, including setting it to the canonic full focus on minimizing statistical error vis-à-vis historical data. While this can be seen as a strength, it is also a limitation until the data science and m/MMM community has built experience and familiarity with this functionality. We hence invite modelers to try out different weights across the objectives and to share their experiences with the Robyn community. In reality, business plausibility is a criterion used in model evaluation as much as or more than statistical fit (Calatrava et al. 2023; Taylor 2023). A model where spend and effect share more closely align will be deemed more plausible than a model that suggests major deviations or produces extreme effects (Manchanda, Rossi, and Chintagunta 2004; Taylor 2023). We hope that formalizing such a preference for non-extreme outcomes in estimation of model candidates rather than informally applying it in model selection can be conducive to effective application.

Another limitation of note is that Robyn, while it accepts other marketing mix actions such as price as "other variables," currently only supports mMM (and not MMM, Arora et al. 2023). This primary focus on mMM derives from market need: small and midsize businesses that thus far relied on deterministic measurement urgently need support with mMM, to mitigate a risk of "flying blind" in advertising analytics and strategy (Runge 2023). In the future, as Robyn, its community, and its users' needs evolve, an extension into full-fledged MMM is certainly possible.

In addition, there is a number of planned future improvements that we briefly want to list here:

- *Release of a Python version of Robyn*
- *Supporting hierarchical modeling*: Hierarchical models such as a regularized mixed effect model that includes both fixed and random effects and uses regularization to prevent overfitting, can be powerful in m/MMM. This type of model is useful for analyzing complex, hierarchical data structures common in retail and omni-channel advertising. We believe that this functionality would be particularly important for adoption in the US market where DMA-level modeling is often the norm.
- *Multi-response modeling*, also known as multiple output regression, can be an important use case, e.g., when there is a funnel of relevant outcome measures, and two or more related dependent variables need to be predicted (e.g., Srinivasan, Vanhuele and Pauwels 2010).

- *Dynamic models with long-term decomposition*: Such models would be particularly useful to understand and model long-term trends and brand effects in a firm's advertising (Kireyev, Pauwels and Gupta 2016).
- *Interactions, synergy effects and multiplicative models*: Today, Robyn is an additive model. Interactions or synergy effects could detect if and when two or more variables work together to produce a greater effect than the sum of their individual effects (Naik and Raman 2003). The presence of such effects is typically addressed with multiplicative models.
- *Inclusion of reach and frequency*: The World Federation of Advertisers (2023) presented its Halo workstream that deems broader reach-based media planning, including exposure modeling (e.g. reach and frequency, impressions, GRPs), a critical component of advertising measurement.

Finally, as noted previously, other than biases from modeling mishaps and managerial interference, the data themselves can reflect social biases inherent in current market behaviors and responses (Lambrecht and Tucker 2019). These data-inherent social biases can be difficult or impossible to detect from computational approaches alone. We encourage marketing data scientists to dedicate time to specifically considering the potential presence of such biases in their datasets and marketing strategies. Lambrecht and Tucker (2019) present an excellent analysis of the issue and highlight avenues for remediation.

**Appendix A: Robyn Workflow and Code Example**

This appendix demonstrates a workflow for Robyn; compare to the more comprehensive workflow and code demo shown on https://github.com/facebookexperimental/Robyn/blob/main/demo/demo.R (Facebook Experimental 2024c) and explained in a demo video (Zhou 2024). The workflow is designed to provide a simplified and concise overview of the steps involved in a complete modeling run of Robyn, on the simulated data available within the Robyn R package "dt_simulated_weekly" (see Figure 4). The workflow comprises input and output, from data loading to model estimation, budget allocation, and model refresh. In the version here, we added starred lines indicating how the workflow steps map to the steps of the mMM modeling chain in Figure 2 and discussed in Section 2.3. Table 1 in the main text presents a mapping between the modeling steps in Figure 2, the workflow steps here, and the computational components of Robyn discussed in Section 3.

**WORKFLOW AND CODE EXAMPLE FOR ROBYN 3.11.1**

\*\*\*\*\*\*\*\*\*\*\*\*\*\*\*\*\*\*\*\*\* **MODELING STEP *INFER* PER FIGURE 2** \*\*\*\*\*\*\*\*\*\*\*\*\*\*\*\*\*\*\*\*\*

STEP 0: SETUP ENVIRONMENT
- **Input/Steps**: Install Robyn and load into environment
- **Output**: Robyn R package is downloaded installed. This step ensures the latest stable (CRAN) or latest experimental version of Robyn is used for compatibility and new features.
- **Example Command**:
```
install.packages("Robyn")  #Install the latest stable version from CRAN
```
```
# remotes::install_github("facebookexperimental/Robyn/R") #Install the latest dev version from GitHub
```
```
library(Robyn)
```

STEP 1: LOAD DATA
- **Input**: Load in own or load the demo dataset `dt_simulated_weekly` & demo holiday `dt_prophet_holidays`
- **Output**: Data loaded into the R environment. In this step, user loads into Robyn all the necessary data for model building

STEP 2: MODEL SPECIFICATION
- **Input**: Dataset, holiday data, and model configuration parameters

- **Output**: InputCollect object containing all model specifications. This sets up the modeling system with all necessary variables and configurations for Robyn MMM analysis
- **Example Command:**

```
InputCollect <- robyn_inputs(
  dt_input = dt_simulated_weekly,
  dt_holidays = dt_prophet_holidays,
  date_var = "DATE", # date format must be "2020-01-01"
  dep_var = "revenue", # there should be only one dependent variable
  dep_var_type = "revenue", # "revenue" (ROI) or "conversion" (CPA)
  prophet_vars = c("trend", "season", "holiday"), # "trend","season", "weekday" & "holiday"
  prophet_country = "DE", # input country code. Check: dt_prophet_holidays
  context_vars = c("competitor_sales_B", "events"), # e.g. competitors, discount, unemployment etc
  paid_media_spends = c("tv_S", "ooh_S", "print_S", "facebook_S", "search_S"), # mandatory input
  paid_media_vars = c("tv_S", "ooh_S", "print_S", "facebook_I", "search_clicks_P"), # mandatory.
  # paid_media_vars must have same order as paid_media_spends. Use media exposure metrics like
  # impressions, GRP etc. If not applicable, use spend instead.
  organic_vars = "newsletter", # marketing activity without media spend
  # factor_vars = c("events"), # force variables in context_vars or organic_vars to be categorical
  window_start = "2016-01-01",
  window_end = "2018-12-31",
  adstock = "geometric" # geometric, weibull_cdf or weibull_pdf.
)
```

STEP 3: DEFINE AND ADD HYPERPARAMETERS
- **Input**: Definitions of model behavior parameters
- **Output**: Configured hyperparameters added to InputCollect. Hyperparameters like adstock and saturation define how marketing inputs decay over time and how they saturate, impacting model accuracy.
- **Example Values**: `facebook_S_thetas = c(0, 0.3)`: Range for the decay rate of Facebook spend, allowing for quick to moderate decay.
- **Note on regularization via ridge regression**: Lambda is the penalty term for regularised regression. Lambda doesn't need manual definition from the users, because it is set to the range of c(0, 1) by default in hyperparameters and will be scaled to the proper altitude with lambda_max and lambda_min_ratio.

******************** **MODELING STEP *CALIBRATE* PER FIGURE 2** ********************

STEP 4: MODEL CALIBRATION / ADD EXPERIMENTAL INPUT
- **Input**: Calibration data to align model outputs with known outcomes
- **Output**: Adjusted model parameters to reflect real-world effects, Calibration helps in adjusting the model based on ground truth data, enhancing predictive accuracy
- **Example Command**:

```
calibration_input <- data.frame(
channel name must in paid_media_vars
channel = c("facebook_S",  "tv_S", "facebook_S+search_S", "newsletter"),
liftStartDate = as.Date(c("2018-05-01", "2018-04-03", "2018-07-01", "2017-12-01")),
```

```
    liftEndDate = as.Date(c("2018-06-10", "2018-06-03", "2018-07-20", "2017-12-31")),
    liftAbs = c(400000, 300000, 700000, 200),
    spend = c(421000, 7100, 350000, 0),
    confidence = c(0.85, 0.8, 0.99, 0.95),
    metric = c("revenue", "revenue", "revenue", "revenue"),
    calibration_scope = c("immediate", "immediate", "immediate", "immediate")
)
InputCollect <- robyn_inputs(InputCollect = InputCollect, calibration_input = calibration_input)
```

STEP 5: BUILD INITIAL MODEL
- **Input**: InputCollect object
- **Output**: OutputModels object containing model results and diagnostics.

```
OutputModels <- robyn_run(
    InputCollect = InputCollect, # feed in all model specification
    cores = NULL, # NULL defaults to (max available - 1)
    iterations = 2000, # 2000 recommended for the dummy dataset
    trials = 5, # 5 recommended for the sample dataset
    ts_validation = TRUE, # 3-way-split time series for NRMSE validation.
    add_penalty_factor = FALSE # Experimental feature. Use with caution.
    optimize_weights = c(1,1,1) # Set weights for multi-objective optimization across NRMSE,
Decomp.RSSD and MAPE.LIFT; defaults to c(1,1,1). New feature. Use with caution.
)
```

```
OutputCollect <- robyn_outputs(
    InputCollect, OutputModels,
    pareto_fronts = "auto", # automatically pick how many pareto-fronts to fill min_candidates (100)
    # min_candidates = 100, # top pareto models for clustering. Default to 100
    # calibration_constraint = 0.1, # range c(0.01, 0.1) & default at 0.1
    csv_out = "pareto", # "pareto", "all", or NULL (for none)
    clusters = TRUE, # Set to TRUE to cluster similar models by ROAS. See ?robyn_clusters
    export = create_files, # this will create files locally
    plot_folder = robyn_directory, # path for plots exports and files creation
    plot_pareto = create_files # Set to FALSE to deactivate plotting and saving model one-pagers
)
```

********************* MODELING STEP *SELECT* PER FIGURE 2 *********************

STEP 6: SELECT AND SAVE THE MODEL
- **Input**: OutputModels object
- **Output**: Selected model saved and export. This steps enables the user to choose the model that best fits their business context and save it for future use.
- **Example Command**:

```r
select_model <- "1_122_7" # Pick one of the models from OutputCollect to proceed
ExportedModel <- robyn_write(InputCollect, OutputCollect, select_model, export = create_files)
print(ExportedModel)
```

********************* **MODELING STEP *PRESCRIBE* PER FIGURE 2** *********************

STEP 7: GET BUDGET ALLOCATION
- **Input**: Selected model and budget scenarios
- **Output**: Budget allocation recommendations. The budget allocator provides insights on how to allocate marketing budget effectively based on model predictions.
- **Example Command**:

```r
scenario = "max_response":  Optimizes for maximum return on investment.
AllocatorCollect1 <- robyn_allocator(
  InputCollect = InputCollect,
  OutputCollect = OutputCollect,
  select_model = select_model,
  # date_range = "all", # Default to "all"
  # total_budget = NULL, # When NULL, default is total spend in date_range
  channel_constr_low = 0.7,
  channel_constr_up = c(1.2, 1.5, 1.5, 1.5, 1.5),
  # channel_constr_multiplier = 3,
  scenario = "max_response",
  export = create_files
)
```

********************* **FURTHER RELEVANT COMPONENTS** *********************

STEP 8: MODEL REFRESH
- **Input**: JSON file of a previously exported model, new data
- **Output**: Updated model based on new data. Regular updates to the model ensure it remains accurate as new data becomes available.
- **Example Command**:

```r
json_file <- "~/Desktop/Robyn_202211211853_init/RobynModel-1_100_6.json"
RobynRefresh <- robyn_refresh(
  json_file = json_file,
  dt_input = dt_simulated_weekly,
  dt_holidays = dt_prophet_holidays,
  refresh_steps = 13,
  refresh_iters = 1000, # 1k is an estimation
  refresh_trials = 1
```

STEP 9: GET MARGINAL RETURNS
- **Input**: Selected model, specific spend levels
- **Output**: Marginal returns for additional spend. Helps in understanding the incremental benefits of additional marketing spend, aiding in decision-making.
- **Example Command:**

```
Response <- robyn_response(
  InputCollect = InputCollect,
  OutputCollect = OutputCollect,
  select_model = select_model,
  metric_name = "facebook_S"
)
Response$plot
```

OPTIONAL: RECREATE PREVIOUS MODELS AND REPLICATE RESULTS
- **Input**: JSON file of a previously exported model
- **Output**: Recreated model and outputs. Ensures consistency and reproducibility by allowing users to reload and rerun models from saved states.

HYPERPARAMETERS EXPLANATION:
- **Geometric Adstock**: Theta values typically range from 0 to 0.8, reflecting the percentage of impact carried over to the next period. For example, a theta of 0.3 for TV means that 30% of the advertising effect remains the next day.
- **Weibull CDF Adstock**: Shape and scale parameters allow for a flexible decay rate. Shape influences the curve's form (S-shaped or L-shaped), while scale adjusts the inflection point of decay. Common bounds are shape (0, 2) and scale (0, 0.1).
- **Weibull PDF Adstock**: Similar to CDF but allows for lagged effects, with shape influencing the peak timing and scale adjusting the peak's position. Recommended bounds are shape (0.0001, 10) and scale (0, 0.1) for varied effects.
- **Hill Saturation Function**: Alpha and gamma parameters define the saturation curve's shape and inflection point. Typical bounds are alpha (0.5, 3) for curve shape and gamma (0.3, 1) for the saturation point.

CALIBRATION:
- **Calibration Input**: Incorporates experimental results to fine-tune the model. For instance, if a Facebook campaign shows a specific lift in revenue during a test period, this data can be used to adjust the model parameters to better reflect this observed effect.
- **Example Values**:
    - `liftAbs = 400000`: Indicates an observed increase in revenue attributed to the campaign.
    - `spend = 421000`: The associated spend during the campaign period.
- **Reasoning**: Calibration ensures the model's predictions align with real-world outcomes, enhancing its predictive accuracy and reliability.